\documentclass[times]{jnmauth}

\usepackage{moreverb}

\newcommand\BibTeX{{\rmfamily B\kern-.05em \textsc{i\kern-.025em b}\kern-.08em
T\kern-.1667em\lower.7ex\hbox{E}\kern-.125emX}}

\def\volumeyear{2016}

\usepackage{siunitx}

\begin{document}

\runningheads{M.~Bonazzoli, V.~Dolean, F.~Rapetti, P.-H.~Tournier}{Parallel preconditioners and high order elements for microwave imaging}

\title{Parallel preconditioners for high order discretizations arising from full system modeling for brain microwave imaging}

\author{Marcella Bonazzoli\affil{1}\corrauth, Victorita Dolean\affil{1,2}, Francesca Rapetti\affil{1}, Pierre-Henri Tournier\affil{3}}

\address{\affilnum{1} Laboratoire J.A.~Dieudonn\'e, University of Nice Sophia Antipolis, Parc Valrose, 06108 Nice Cedex 02, France. 
\break E-mail: {marcella.bonazzoli@unice.fr}, {victorita.dolean@unice.fr}, {francesca.rapetti@unice.fr}
\break \affilnum{2} Department of Mathematics and Statistics, University of Strathclyde, Glasgow, UK.  
\break E-mail: {victorita.dolean@strath.ac.uk}
\break \affilnum{3} INRIA Paris, Alpines, and UPMC - Univ Paris 6, CNRS UMR 7598, Laboratoire Jacques-Louis
Lions, France.
\break E-mail: {pierre-henri.tournier@inria.fr}}

\corraddr{Laboratoire J.A.~Dieudonn\'e, University of Nice Sophia Antipolis, Parc Valrose, 06108 Nice Cedex 02, France. E-mail: marcella.bonazzoli@unice.fr.}

\cgsn{French National Research Agency (ANR), project MEDIMAX}{ANR-13-MONU-0012.}

\begin{abstract}
This paper combines the use of high order finite element methods with parallel preconditioners of domain decomposition type for solving electromagnetic problems arising from brain microwave imaging. 
The numerical algorithms involved in such complex imaging systems are computationally expensive since they require solving the direct problem of Maxwell's equations several times. Moreover, wave propagation problems in the high frequency regime are challenging because a sufficiently high number of unknowns is required to accurately represent the solution.
In order to use these algorithms in practice for brain stroke diagnosis, running time should be reasonable.
The method presented in this paper, coupling high order finite elements and parallel preconditioners, makes it possible to reduce the overall computational cost and simulation time while maintaining accuracy. \\
\end{abstract}


\maketitle

\section{Introduction}

The context of this work is the solution of an inverse problem associated with the time-harmonic Maxwell's equations, with the aim of estimating the dielectric properties of the brain tissues of a patient affected by a brain stroke. 
Strokes can be cast in two major categories, ischemic (80\% of strokes) and hemorrhagic (20\% of strokes), which result in opposite variations of these dielectric properties. In the following, we briefly describe this particular medical context as well as the application motivating the numerical model.

During an ischemic stroke the blood supply to a part of the brain is interrupted by the formation of a blood clot inside a vessel, while a hemorrhagic stroke occurs when a blood vessel bursts inside the brain. 
It is essential to determine the type of stroke in the shortest possible time in order to start the correct treatment, which is opposite in the two situations: in the first case the blood flow should be restored, while in the second one we need to lower the blood pressure.
Note that it is vital to make a clear distinction between the two types of stroke before treating the patient: the treatment that suits an ischemic stroke would be fatal if applied to a hemorrhagic stroke and vice versa. Moreover, it is desirable to be able to monitor continuously the effect of the treatment on the evolution of the stroke during the hospitalization.

Usually stroke diagnosis relies mainly on two types of imaging techniques: MRI (magnetic resonance imaging) or CT scan (computerized tomography scan). These are very precise techniques, especially the MRI with a spatial resolution of $1$\,mm. However, a MRI machine is too big to be carried in ambulance vehicles and it is too expensive; a CT scan, which consists in measuring the absorption of X-rays by the brain, is harmful and cannot be used to monitor continuously the patient in hospital. 

A novel competitive technique with these traditional imaging modalities is microwave tomography.
With microwave imaging in a range of frequencies between $100$\,MHz and several GHz, the tissues are well differentiated and they can be imaged on the basis of their dielectric properties. The electromagnetic emissions are lower than the ones from mobile phones and the spatial resolution is good ($5-7$\,mm). 
The first works on microwave imaging date back to 1989 when Lin and Clarke tested experimentally the detection of cerebral edema (excessive accumulation of water in the brain) using a frequency signal of $2.4$\,GHz in a head phantom. Other works followed, but almost always on phantoms or synthetic simplified models \cite{Semenov:2008:MTB}. Despite these encouraging results, there is still no microwave device for medical diagnosis. 
The techniques designed by the University of Chalmers (Gothenburg, Sweden) \cite{Persson:2014:MBS} and by EMTensor GmbH (Vienna, Austria) \cite{Semenov:2014:ETB} rely on technologies and softwares developed only in recent years. In both cases the improvement in terms of reliability, price and miniaturization of electromagnetic sensors is a key factor. 
In this approach, it is necessary to transfer the data to a remote HPC machine. The rapid telephony standards such as 4G and 5G allow to send the acquired measurements of the patient's brain to a supercomputer that will compute the 3D images. Then these images can be quickly transmitted from the computer to the hospital by ADSL or fiber network.

\begin{figure}
\centering 
\includegraphics[width=0.49\textwidth]{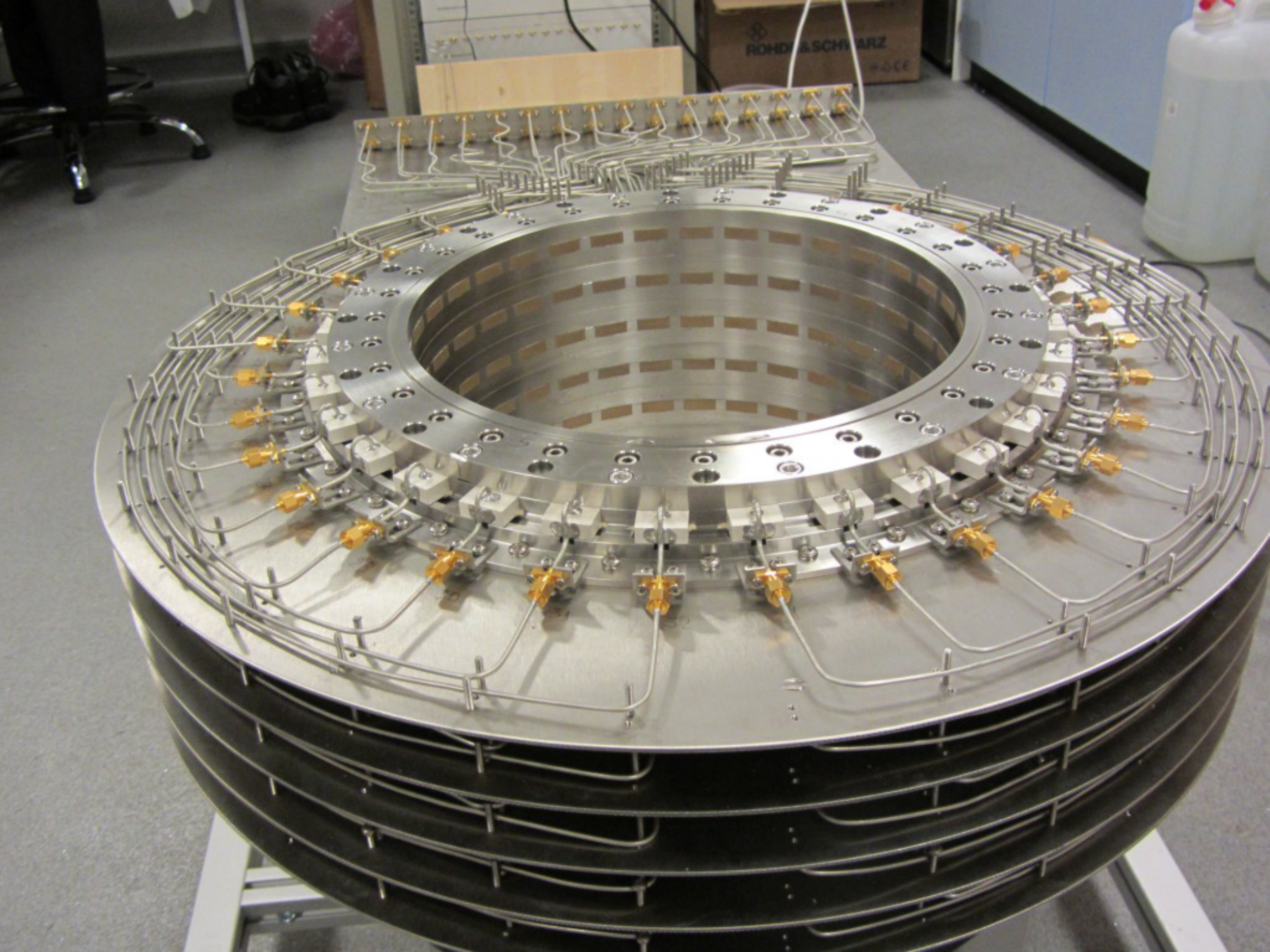} \quad 
\includegraphics[width=0.39\textwidth]{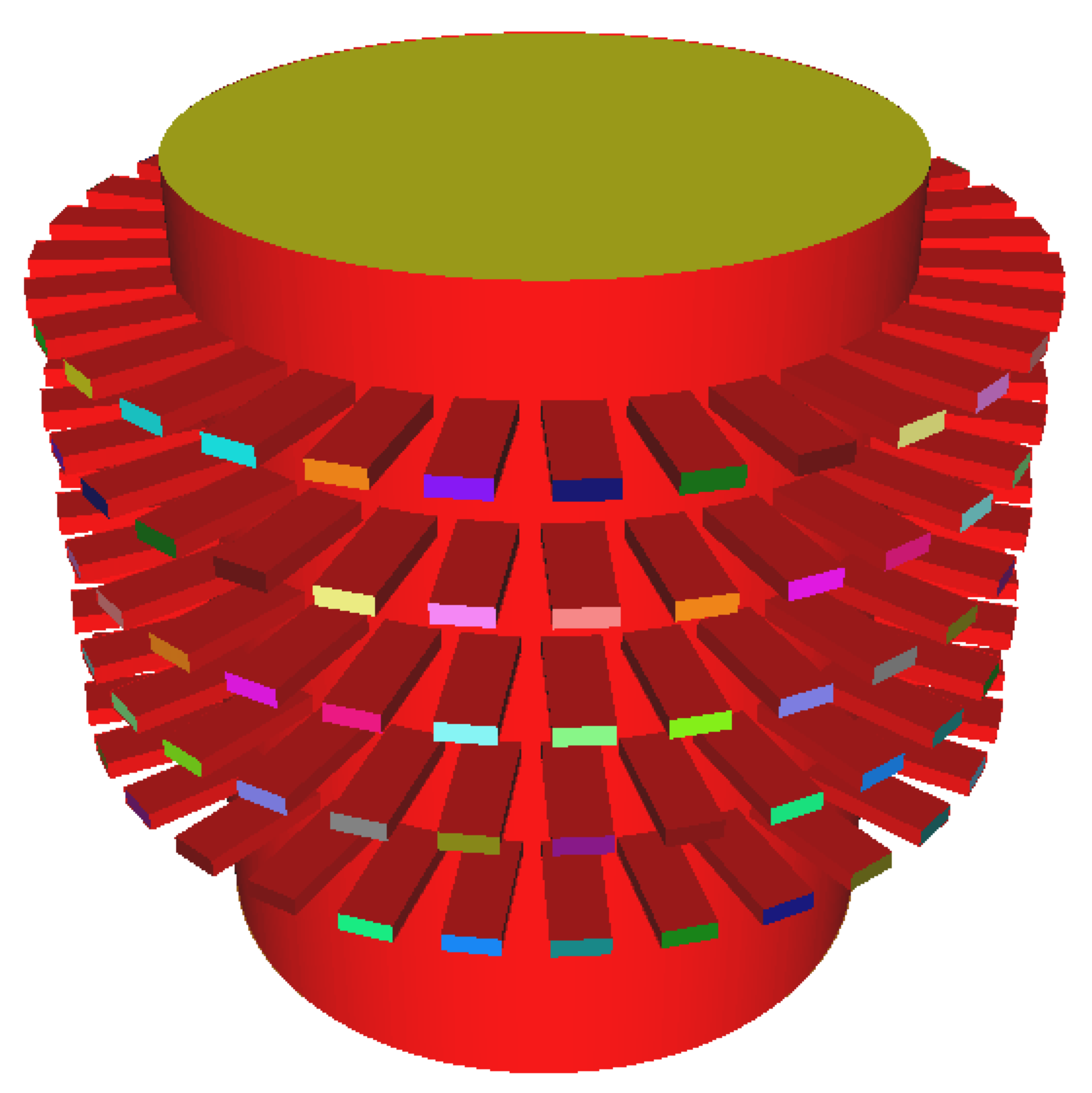} 
\caption{Imaging chamber of EMTensor (no copyright infringement intended).} 
\label{fig:chamber}
\end{figure}
Figure~\ref{fig:chamber} shows the initial microwave imaging system prototype of EMTensor: it is composed of $5$ rings of $32$ ceramic-loaded rectangular waveguides around a metallic cylindrical chamber of diameter $28.5$\,cm and total height $28$\,cm, into which the patient head is inserted. Each of the $160$ antennas alternately transmits a signal at a fixed frequency, typically $1$\,GHz. The electromagnetic wave propagates inside the chamber and in the object to be imaged according to its electromagnetic properties. The retrieved data then consist in the reflection and transmission coefficients measured by the $160$ receiving antennas, which are  
used as input for the inverse problem.
Since the inversion loop requires to solve repeatedly the \emph{direct} problem of the time-harmonic Maxwell's equations in high frequency regime, an accurate and fast solver of the direct problem is needed.
In this paper accuracy is provided by a high order edge finite element discretization, and the resulting linear system is solved efficiently with the iterative method GMRES preconditioned with a parallel preconditioner based on domain decomposition methods.

The paper is organized as follows. In Section \ref{sec:mathmodel} the mathematical model of time-harmonic Maxwell's equations in curl-curl form is presented, together with the associated boundary value problem to solve. In Section \ref{sec:edgelements} the discretization method using high order edge finite elements is briefly described and in Section \ref{sec:domaindecomposition} the parallel preconditioner based on domain decomposition is introduced. Section \ref{sec:numericalresults} contains in the first part a comparison with experimental measurements; in the second part we assess the efficiency of high order edge finite elements compared to the standard lowest order edge elements in terms of accuracy and computing time.

\section{Mathematical model}
\label{sec:mathmodel}

To work in the frequency domain, we assume that the electric field $\mathcal{E} (\mathbf{x}, t) = \mathrm{Re}(\mathbf{E}(\mathbf{x}) e^{\mathtt{i}\omega t})$ has harmonic dependence on time of angular frequency $\omega$, where $\mathbf{E}$ is its complex amplitude depending only on the space variable $\mathbf{x}$. Thus, considering a non magnetic medium with magnetic permeability $\mu$ equal to the free space magnetic permeability $\mu_0$, we can get the following second order time-harmonic Maxwell's equation:
\begin{equation}\label{eq:MaxwellSecondOrder}
\nabla\times\left(\nabla\times \mathbf{E}\right) - {\gamma^2} \mathbf{E} = {\bf 0},
\quad \gamma = \omega \sqrt{\mu \varepsilon_{\sigma}},  \quad 
\varepsilon_{\sigma} = \varepsilon - {\tt i} \frac{\sigma}{\omega}.
\end{equation}
Here $\varepsilon_{\sigma}$ is the complex valued electric permittivity, related to the dissipation-free electric permittivity $\varepsilon$ and to the electrical conductivity $\sigma$ of the medium.
Notice that if $\sigma=0$, we have $\gamma = \tilde\omega$, $\tilde\omega = \omega\sqrt{\mu \varepsilon}$ being the wavenumber.
Equation~\eqref{eq:MaxwellSecondOrder} is solved in the computational domain $\Omega \subset \mathbb{R}^3$ shown in Figure~\ref{fig:chamber} (right), with metallic boundary conditions
\begin{equation}
\label{eq:metallicbc}
\mathbf{E}\times\mathbf{n} = \mathbf{0} \text{ on } \Gamma_{\text{w}},
\end{equation}
on the cylinder and waveguides walls $\Gamma_{\text{w}}$,
and with impedance boundary conditions on the port $\Gamma_{j}$ of the $j$-th waveguide, which transmits the signal, and on the ports $\Gamma_{i}$ of the receiving waveguides, $i = 1,\dots,160, i \neq j$:
\begin{alignat}{2}
(\nabla\times\mathbf{E})\times\mathbf{n} + {\tt i} \beta \mathbf{n} \times 
(\mathbf{E} \times \mathbf{n}) &=\mathbf{g}_{j} & &\text{ on } \Gamma_{j} , \label{eqwgj} \\
(\nabla\times\mathbf{E})\times\mathbf{n} + {\tt i} \beta \mathbf{n} \times
(\mathbf{E} \times \mathbf{n}) &= \mathbf{0} & &\text{ on } \Gamma_{i} \mbox{ , } i \neq j. \label{eqwgi}
\end{alignat}
Here $\mathbf{n}$ is the unit outward normal to $\partial \Omega$ and $\beta \in  \mathbb{R}_{>0}$ is the propagation wavenumber along the waveguides. Equation (\ref{eqwgj}) imposes an incident wave which corresponds to the excitation of the TE$_{10}$ fundamental mode  $\mathbf{E}_j^0$ of the  $j$-th waveguide, with $\mathbf{g}_{j} = (\nabla \times \mathbf{E}_j^0) \times \mathbf{n} +\mathtt{i} \beta \mathbf{n} \times  (\mathbf{E}_j^0 \times \mathbf{n})$. Equation (\ref{eqwgi}) is an absorbing boundary condition of Silver-M{\"u}ller giving a first order approximation of a transparent boundary condition on the outer port of the receiving waveguides $i = 1,\dots,160,$ with $i \neq j$. The bottom of the chamber is considered metallic, and we impose an impedance boundary condition on the top of the chamber.

The variational formulation corresponding to equation~\eqref{eq:MaxwellSecondOrder} together with boundary conditions~\eqref{eq:metallicbc},~\eqref{eqwgj},~\eqref{eqwgi} is: find $\mathbf{E} \in V$ such that
\begin{multline*}
\int_{\Omega} \Bigl[
(\nabla\times \mathbf{E})\cdot (\nabla\times \mathbf{v})-\gamma^2\mathbf{E}\cdot \mathbf{v}  \Bigr] 
+ \int_{\bigcup_{i=1}^{160}\Gamma_i} {\tt i} \beta (\mathbf{E} \times \mathbf{n}) \cdot
(\mathbf{v} \times \mathbf{n} ) 
= \int_{\Gamma_{j}} \mathbf{g}_j\cdot \mathbf{v} 
\quad \forall \mathbf{v}\in V,
\end{multline*}
with $V=\{\mathbf{v}\in H(\text{curl},\Omega), \mathbf{v} \times \mathbf{n} = 0\text{ on }\Gamma_{\text{w}}\}$, 
where $H(\text{curl},\Omega)$ is the space of square integrable functions whose curl is also square integrable.
Note that $\mathbf{g}_j$ depends on which waveguide transmits the signal and this corresponds to a different right-hand side of the linear system resulting from the finite element discretization. On the other hand, the matrix of the linear system is the same for every transmitting waveguide.

\section{High order edge finite elements}
\label{sec:edgelements}

To write a finite element discretization of the variational problem we introduce a tetrahedral mesh $\mathcal{T}_h$ of the domain $\Omega$ and a finite dimensional subspace $V_h\subset H(\text{curl},\Omega)$.
The simplest possible conformal discretization for the space $H(\text{curl},\Omega)$ is given by the low order \emph{N\'ed\'elec edge finite elements} (of polynomial degree $r=1$) \cite{Nedelec:1980:MFE}: 
for a tetrahedron $T \in \mathcal{T}_h$, the local basis functions are associated with the oriented edges $e=\{n_i,n_j\}$ of $T$ as follows
\[
\mathbf{w}^e = \lambda_{i}\nabla \lambda_{j} - \lambda_{j}\nabla\lambda_{i},
\]
where the $\lambda_{\ell}$ are the barycentric coordinates of a point with respect to the node $n_\ell$. It can be shown that edge finite elements guarantee the continuity of the tangential component across faces shared by adjacent tetrahedra, they thus fit the continuity properties of the electric field.

The finite element discretization is obtained by writing the discretized field over each tetrahedron $T$ as $\mathbf{E}_h= \sum_{e \in T} c_e {\bf w}^e$, a linear combination with coefficients $c_e$ of the basis functions associated with the edges $e$ of $T$, and the coefficients $c_e$ will be the unknowns of the resulting linear system.
For edge finite elements of degree $1$ these coefficients can be interpreted as the \emph{circulations} of ${\bf E}_h$ along the edges of the tetrahedra:
\[ 
c_e = \frac{1}{|e|} \int_e {\bf E}_h \cdot {\bf t}_e,
\]
where ${\bf t}_e$ is the tangent vector to the edge $e$ of length $|e|$, the length of $e$. This is a consequence of the fact that the basis functions are in duality with the degrees of freedom given by the circulations, that is:
\[
\frac{1}{|e|} \int_e \mathbf{w}^{e'} \cdot {\bf t}_e = 
\begin{cases}
1 \quad \text{if } e=e', \\
0 \quad \text{if } e \neq e'.
\end{cases}
\]


In order to have a higher numerical accuracy with the same total number of unknowns, we consider a \emph{high order} edge element discretization, choosing the high order extension of N\'ed\'elec elements presented in \cite{Rapetti:2007:HOE} and \cite{RapBos:2009:WFH}. The definition of the basis functions is rather simple since it only involves the barycentric coordinates of the tetrahedron.
Given a multi-index $\mathbf{k} = (k_1, k_2, k_3, k_4)$ of weight $k = k_1+k_2+k_3+k_4$ (where the $k_i, i=1,2,3,4,$ are non negative integers), we denote by $\lambda^\mathbf{k}$ the product $\lambda_1^{k_1}\lambda_2^{k_2}\lambda_3^{k_3}\lambda_4^{k_4}$.
The local generators of polynomial degree $r=k+1$ ($k\ge0$) over the tetrahedron $T$ are defined as
\[
\mathbf{w}^{\{\mathbf{k},e\}} = \lambda^\mathbf{k} \mathbf{w}^e,
\]
for all edges $e$ of the tetrahedron $T$, and for all multi-indices $\mathbf{k}$ of weight $k$. 
Note that these high order elements still yield a conformal discretization of $H(\text{curl},\Omega)$: indeed, they are products between the degree $1$ N\'ed\'elec elements $\mathbf{w}^e$, which are curl-conforming, and the continuous functions $\lambda^\mathbf{k}$.
However, some of these high order generators ($r>1$) are linearly dependent: the selection of a linearly independent subset to constitute an actual basis is described in \cite{BDHR:2016:OSP}, which provides further details about the implementation of these finite elements. 
Moreover, the duality property, which is practical for the implementation, is not satisfied for high order generators, but it can be easily restored as explained in \cite{BonRap:2015:duality}.

Duality is needed for instance in FreeFem++, an open source domain specific language (DSL) specialized for solving boundary value problems by using variational discretizations (finite elements, discontinuous Galerkin, hybrid methods, \dots) \cite{Hecht:2012:NDF}. 
Several finite element spaces are available in FreeFem++, and the user can also add new finite elements, provided that the duality property is satisfied. For instance we implemented the edge elements in 3d of degree $2$ and $3$, which can be used by loading the plugin {\tt "Element\_Mixte3d"} and declaring the finite element space {\tt fespace} using the keywords {\tt Edge13d}, {\tt Edge23d} respectively (the standard edge elements of degree $1$ were already present in FreeFem++ and thery are called {\tt Edge03d}).

\section{Domain decomposition preconditioning}
\label{sec:domaindecomposition}

\begin{figure}
\centering 
\includegraphics[width=0.39\textwidth]{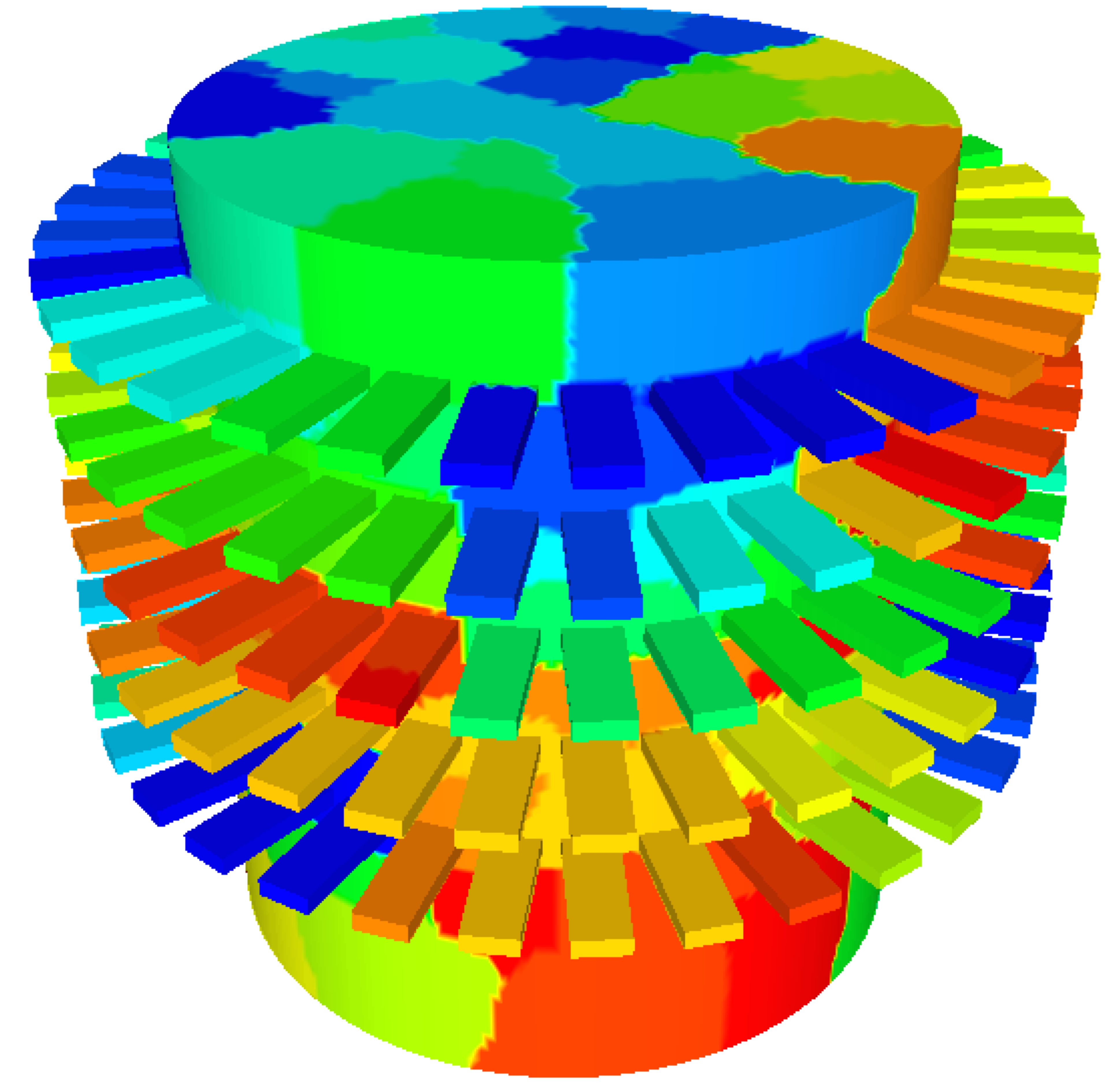} 
\caption{The decomposition of the computational domain into $128$ subdomains.} 
\label{fig:decomp}
\end{figure}

The discretization of the problem presented in Section~\ref{sec:mathmodel} using the high order edge finite elements described in Section~\ref{sec:edgelements} produces a linear system $A {\bf u}_j = {\bf b}_j$ for each transmitting antenna $j$. Direct solvers are not suited for such large linear systems arising from complex three dimensional models because of their high memory cost. On the other hand, matrices resulting from high order discretizations are ill conditioned as shown numerically in \cite{Rapetti:2007:HOE} for similar problems, and preconditioning becomes necessary when using iterative solvers.


Domain decomposition preconditioners are naturally suited to parallel computing and make it possible to deal with smaller subproblems \cite{Dolean:2015:IDD}. 
The domain decomposition preconditioner we employ is called \emph{Optimized Restricted Additive Schwarz} (ORAS):
\[
M^{-1}_{\text{ORAS}} = \sum_{s=1}^{N_{\text{sub}}} R^T_s D_s A_s^{-1}R_s,
\]
where $N_{\text{sub}}$ is the number of overlapping subdomains $\Omega_s$ into which the domain $\Omega$ is decomposed (see Figure~\ref{fig:decomp}).
Here, the matrices $A_s$ are the local matrices of the \emph{subproblems} with impedance boundary conditions $(\nabla\times\mathbf{E})\times\mathbf{n} + {\tt i} \tilde \omega \mathbf{n} \times 
(\mathbf{E} \times \mathbf{n})$ as transmission conditions at the interfaces between subdomains. 
This preconditioner is an extension of the restricted additive Schwarz method proposed by Cai and Sarkis \cite{CaiSar:1999:RAS}, but with more efficient transmission conditions between subdomains than Dirichlet conditions (see for example~\cite{DolGanGer:2009:OSM}).

In order to describe the matrices $R_s, D_s$, let $\mathcal{N}$ be an ordered set of the unknowns of the whole domain and let $\mathcal{N} = \bigcup_{s=1}^{N_{\text{sub}}}\mathcal{N}_s$ be its decomposition into the (non disjoint) ordered subsets corresponding to the different (overlapping) subdomains $\Omega_s$.
The matrix $R_s$ is the restriction matrix from $\Omega$ to the subdomain $\Omega_s$: it is a $\#\mathcal{N}_s \times \#\mathcal{N}$ Boolean matrix and its $(i,j)$ entry is equal to $1$ if the $i$-th unknown in $\mathcal{N}_s$ is the $j$-th one in $\mathcal{N}$. Notice that $R^T_s$ is then the extension matrix from the subdomain $\Omega_s$ to $\Omega$. 
The matrix $D_s$ is a $\#\mathcal{N}_s \times \#\mathcal{N}_s$ diagonal matrix that gives a discrete partition of unity, i.e. $\sum_{s=1}^{N_{\text{sub}}} R^T_s D_s R_s = I$; in particular the matrices $D_s$ deal with the unknowns that belong to the overlap between subdomains. 

The preconditioner without the partition of unity matrices $D_s$, $M^{-1}_{\text{OAS}} = \sum_{s=1}^{N_{\text{sub}}} R^T_s A_s^{-1}R_s$, which is called Optimized Additive Schwarz (OAS), would be symmetric for symmetric problems, but in practice it gives a slower convergence with respect to $M^{-1}_{\text{ORAS}}$, as shown for instance in \cite{BDHR:2016:OSP}.

These domain decomposition preconditioners are implemented in the library HPDDM \cite{JolHecNat:2013:hpddm}, an open source high-performance unified framework for domain decomposition methods. HPDDM can be interfaced with various programming languages and open source finite element libraries such as  FreeFem++, which we use in the simulations.

\section{Numerical results}
\label{sec:numericalresults}

In this section, all linear systems resulting from the edge finite elements discretizations are solved by GMRES preconditioned with the ORAS preconditioner as implemented in HPDDM. Each linear system to solve has several right-hand sides (one per transmitter), and we use a pseudo-block method implemented inside GMRES which consists in fusing the multiple arithmetic operations corresponding to each right-hand side (matrix-vector products, dot products) in order to achieve higher arithmetic intensity.

All the simulations are performed in FreeFem++ interfaced with HPDDM. Results were obtained on the Curie supercomputer (GENCI-CEA).



In the following subsections, we first validate our numerical modeling of the imaging chamber by comparing the results of the simulation with experimental measurements obtained by EMTensor. Then, we illustrate the efficiency of the high order finite elements presented in Section~\ref{sec:edgelements} over the classical lowest order ones in terms of running time and accuracy.

\subsection{Comparison with experimental measurements}

The physical quantity that can be acquired by the measurement system  of the imaging chamber shown in Figure~\ref{fig:chamber} is the scattering matrix ($S$ matrix), which gathers the complex \emph{reflection and transmission coefficients} measured by the $160$ receiving antennas for a signal transmitted by one of these $160$ antennas successively. A set of measurements then consists in a complex matrix of size $160 \times 160$. In order to compute the numerical counterparts of these reflection and transmission coefficients, we use the following formula, which is appropriate in the case of open waveguides:

\begin{equation}
\label{sij}
S_{ij} = \frac{\int_{\Gamma_i} \overline{\mathbf{E}_j} \cdot \mathbf{E}_i^0}{\int_{\Gamma_i} |\mathbf{E}_i^0|^2},  \quad i, j = 1,\dots,160, 
\end{equation}
where $\mathbf{E}_j$ is the solution of the problem where the $j$-th waveguide transmits the signal, and $\mathbf{E}_i^0$ is the TE$_{10}$ fundamental mode of the $i$-th receiving waveguide ($\overline{\mathbf{E}_j}$ denotes the complex conjugate of $\mathbf{E}_j$).
The $S_{ij}$ with $i \neq j$ are the transmission coefficients, and the $S_{jj}$ are the reflection coefficients.

For this comparison of the computed coefficients with the measured ones, the imaging chamber is filled with a homogenous matching solution. The electric permittivity $\varepsilon$ of the matching solution is chosen by EMTensor in order to minimize contrasts with the ceramic-loaded waveguides and with the different brain tissues. The choice of the conductivity $\sigma$ of the matching solution is a compromise between the minimization of reflection artifacts from metallic boundaries and the desire to have best possible signal-to-noise ratio. Here the relative complex permittivity of the matching solution at frequency $f = $ \SI{1}{\giga\hertz} is $\varepsilon_r^{\text{gel}} = 44 - 20\mathtt{i}$. The relative complex permittivity inside the ceramic-loaded waveguides is $\varepsilon_r^{\text{cer}} = 59 - 0\mathtt{i}$. Here with $\varepsilon_r$ we mean the ratio between the complex permittivity $\varepsilon_\sigma$ and the permittivity of free space $\varepsilon_0$.

\begin{figure}
\centering 
\includegraphics[width=0.7\textwidth]{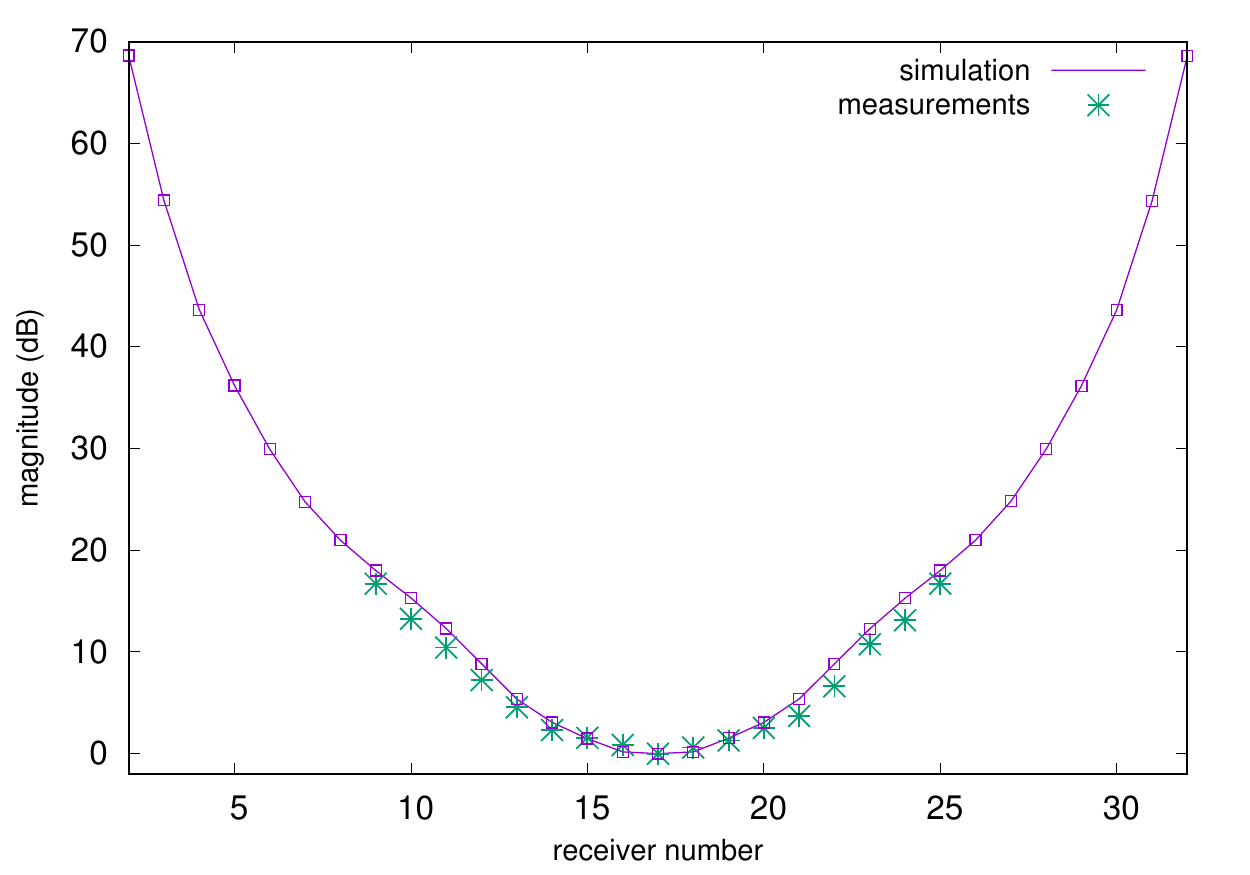} \\
\includegraphics[width=0.71\textwidth]{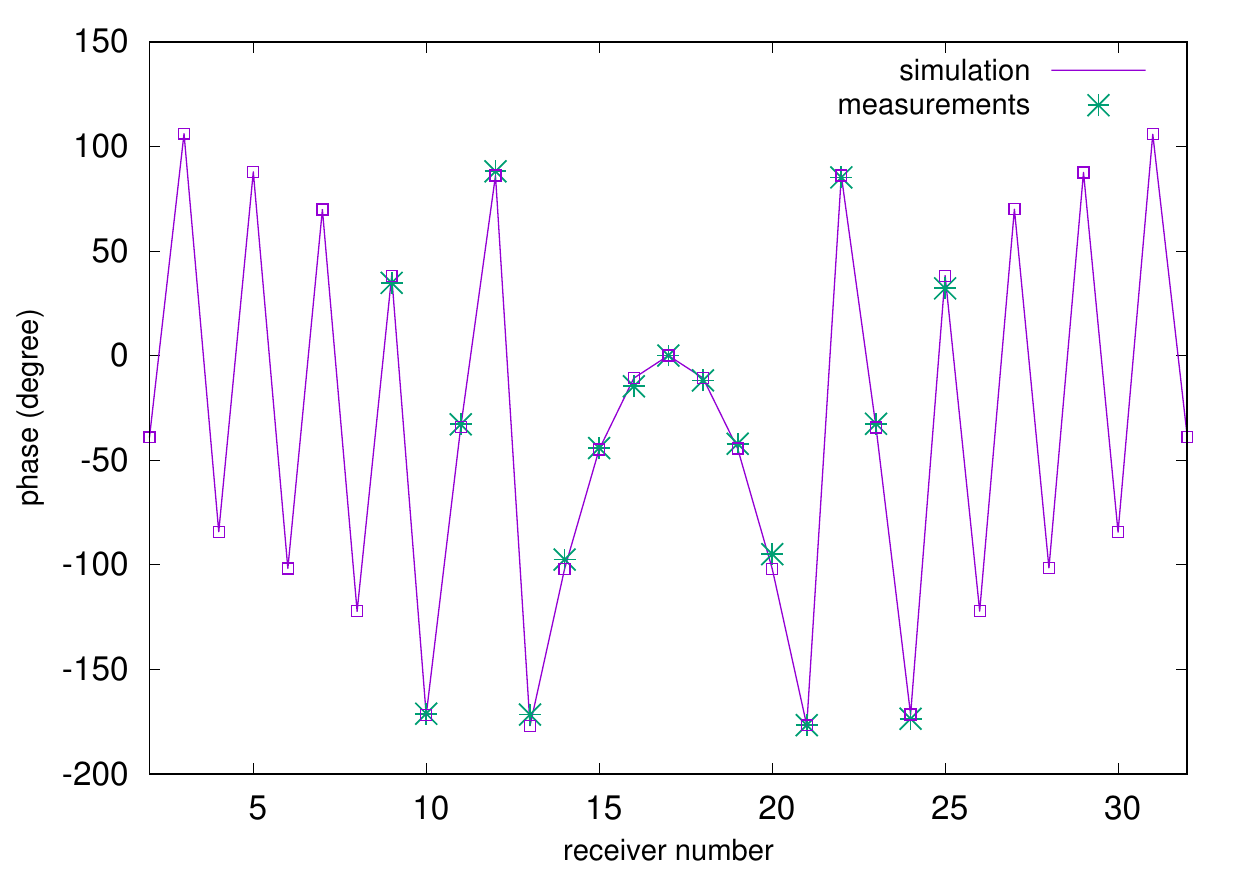} 
\caption{The normalized magnitude (top) and phase (bottom) of the transmission coefficients computed with the simulation and measured experimentally.} 
\label{fig:compexp}
\end{figure}
For this test case, the set of experimental data given by EMTensor consists in transmission coefficients for transmitting antennas in the second ring from the top.
Figure~\ref{fig:compexp} shows the normalized magnitude (dB) and phase (degree) of the complex coefficients $S_{ij}$ corresponding to a transmitting antenna in the second ring from the top and to the $31$ receiving antennas in the middle ring (notice that measured coefficients are available only for $17$ receiving antennas). The magnitude in dB is calculated as $20 \log_{10} (|S_{ij} |)$.  
The computed coefficients are obtained by solving the direct problem with edge finite elements of polynomial degree $r=2$.
We can see that the computed transmission coefficients are in very good agreement with the measurements.


\subsection{Efficiency of high order finite elements}

\begin{figure}
\centering 
\includegraphics[width=0.49\textwidth]{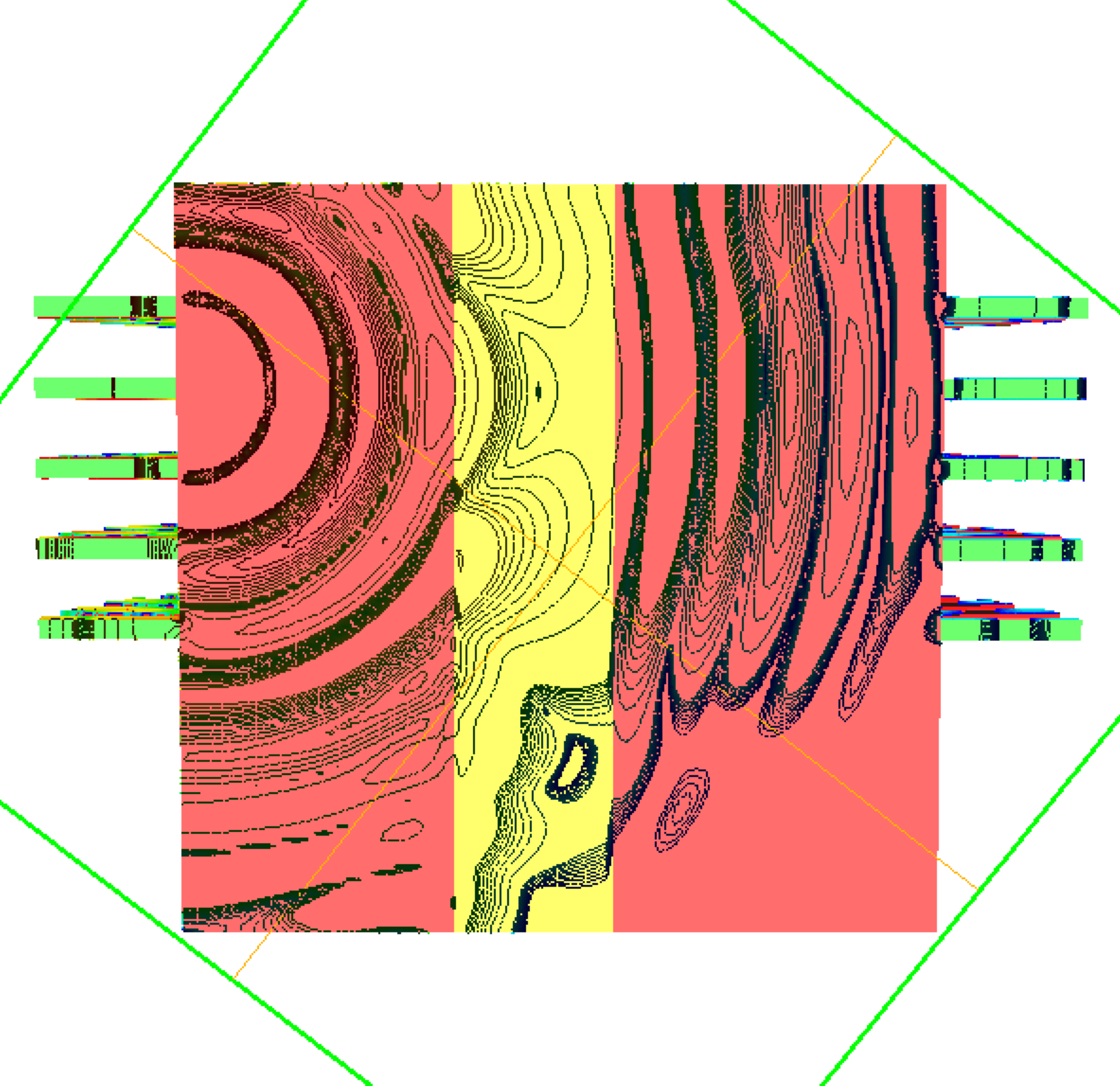}
\caption{Slice of the imaging chamber, showing the non-dissipative plastic-filled cylinder and some isolines of the norm of the real part of the total field $\mathbf{E}$.} 
\label{fig:chambertube}
\end{figure}
The goal of the following numerical experiments is to assess the efficiency of the high order finite elements described in Section~\ref{sec:edgelements} compared to the classical lowest order edge elements in terms of accuracy and computing time, which are of great importance for such an application in brain imaging. For this test case, a non-dissipative plastic-filled cylinder of diameter \SI{6}{\centi\meter} and relative permittivity $\varepsilon_r^{\text{cyl}} = 3$ is inserted in the imaging chamber and surrounded by matching solution of relative complex permittivity $\varepsilon_r^{\text{gel}} = 44 - 20\mathtt{i}$ (see Figure~\ref{fig:chambertube}). We consider the $32$ antennas of the second ring from the top as transmitting antennas at frequency $f = \SI{1}{\giga\hertz}$, and all $160$ antennas are receiving. We evaluate the relative error on the reflection and transmission coefficients $S_{ij}$ with respect to the coefficients $S_{ij}^{\text{ref}}$ computed from a reference solution. The relative error is calculated with the following formula:
\begin{equation}
\label{errsij}
E = \frac{\sqrt{\sum_{j,i} | S_{ij} - S_{ij}^{\text{ref}}|^2}}{\sqrt{\sum_{j,i} |S_{ij}^{\text{ref}}|^2}}.
\end{equation}

\begin{figure}
\centering 
\includegraphics[width=0.49\textwidth]{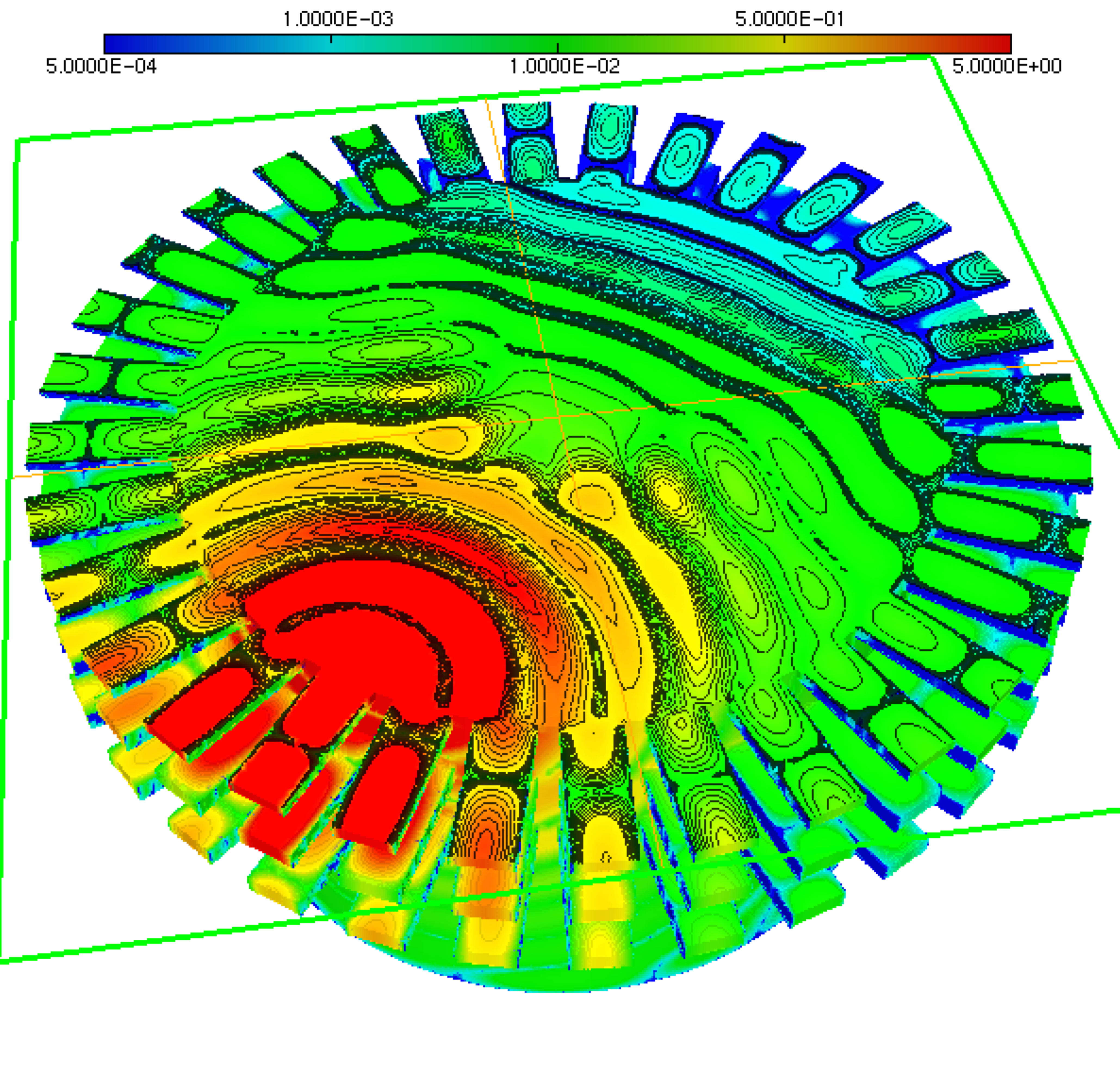} \;
\includegraphics[width=0.49\textwidth]{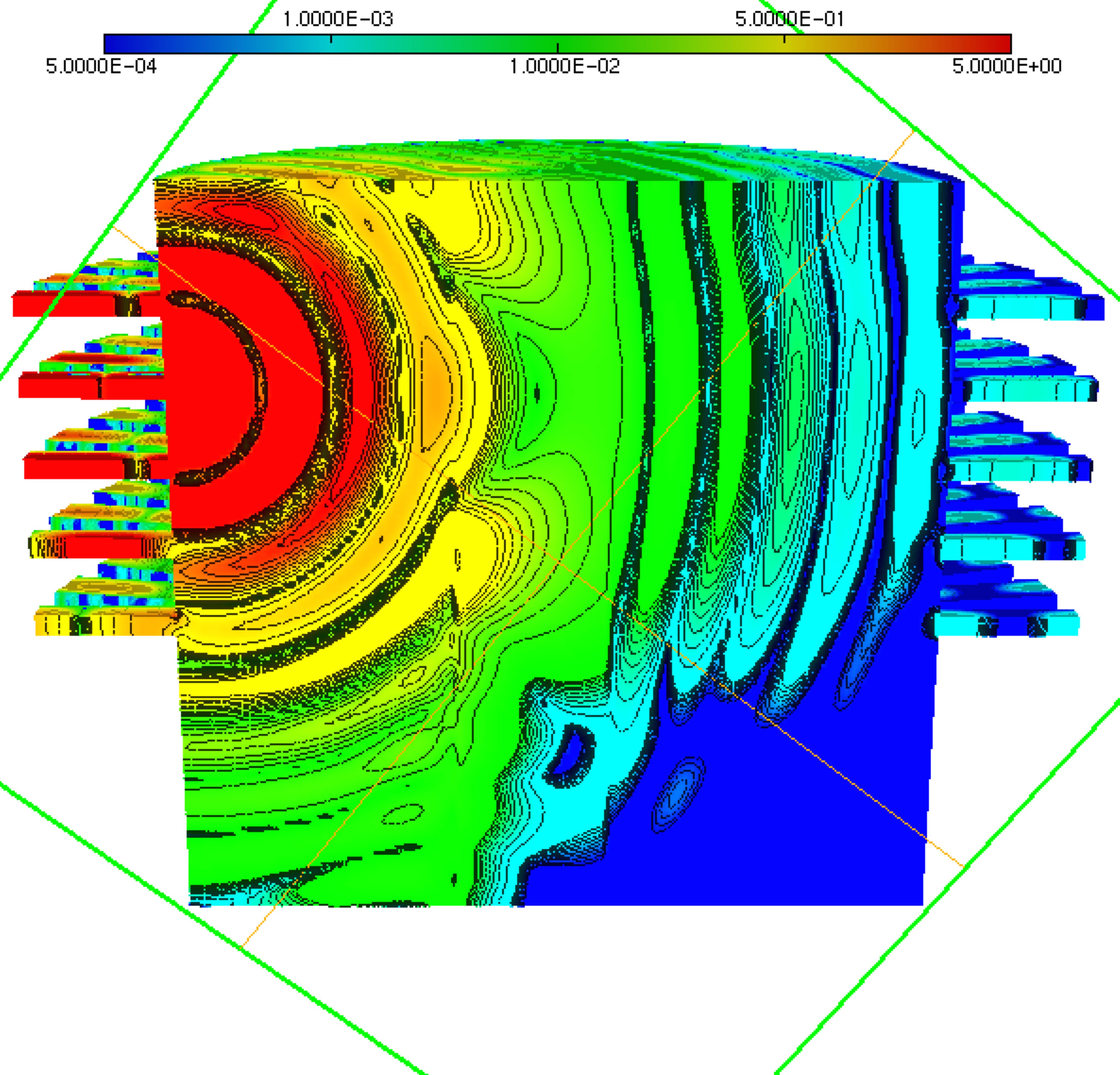} 
\caption{Slices showing the norm of the real part of the total field $\mathbf{E}$ in the imaging chamber with the plastic-filled cylinder inside, for a transmitting antenna in the second ring from the top.} 
\label{fig:solutiontube}
\end{figure}
The reference solution is computed on a fine mesh of approximately $18$ million tetrahedra using edge finite elements of degree $r=2$, resulting in $114$ million unknowns. Slices in Figures~\ref{fig:chambertube} and~\ref{fig:solutiontube} show the computational domain and the solution $\mathbf{E}$ for one transmitting antenna in the second ring from the top.

We compare the computing time and the relative error~\eqref{errsij} for different numbers of unknowns corresponding to several mesh sizes, for approximation degrees $r=1$ and $r=2$. All these simulations are done using $512$ subdomains with one MPI process and two OpenMP threads per subdomain, for a total of $1024$ cores on the Curie supercomputer.

\begin{figure}
\centering 
\includegraphics[width=0.7\textwidth]{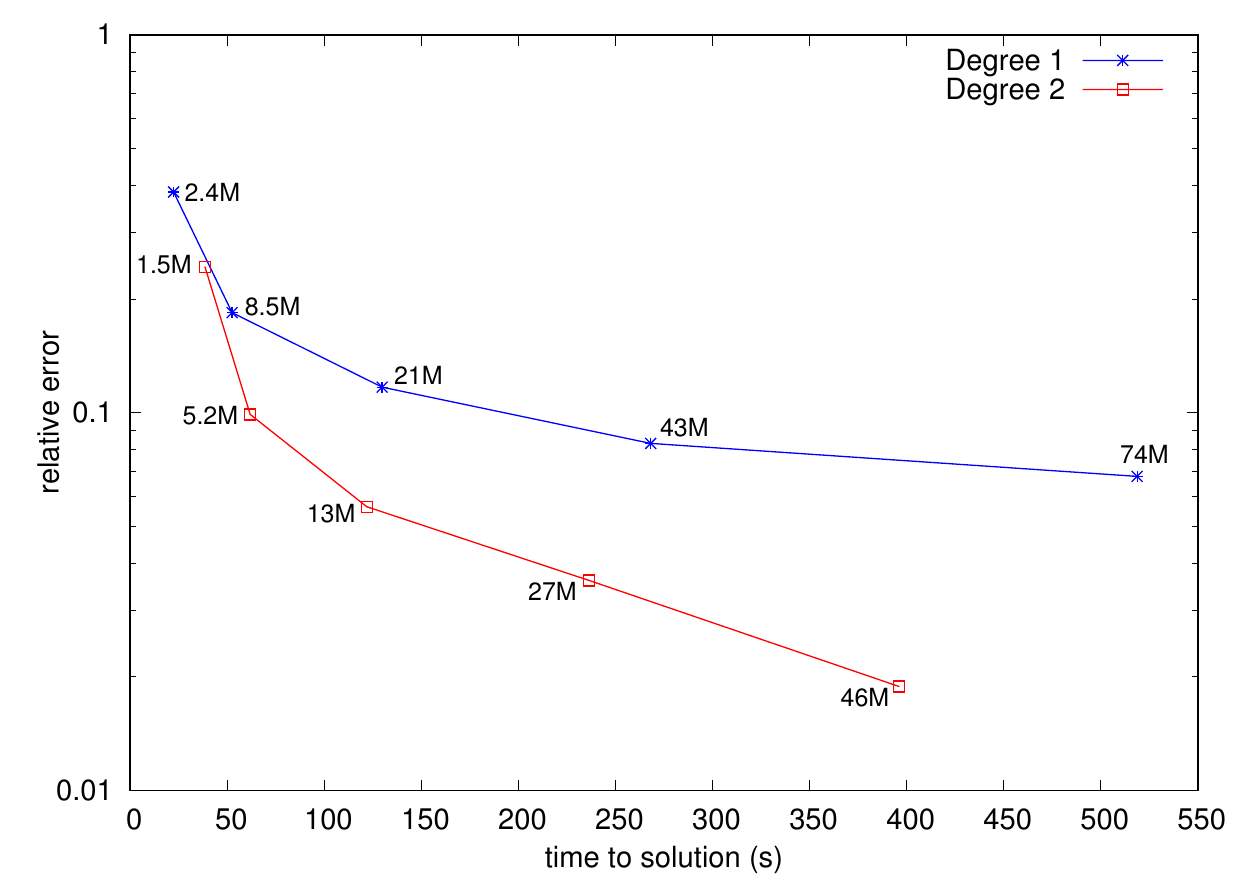}
\caption{Time to solution (seconds) and relative error on the computed $S_{ij}$ with respect to the reference solution, using edge finite elements of degree $1$ and degree $2$ for different mesh sizes. The total number of unknowns in millions is also reported for each simulation.} 
\label{fig:compho}
\end{figure}

We report the results in Table~\ref{tab:degree12} and in Figure~\ref{fig:compho}. As we can see, the high order approximation ($r = 2$) allows to attain a given accuracy with much fewer unknowns and much less computing time than the lowest order approximation ($r = 1$). For example, at a given accuracy of $E \approx 0.1$, the finite element discretization of degree $r = 1$ requires $21$ million unknowns and a computing time of $130$ seconds, while the high order finite element discretization ($r = 2$) only needs $5$ million unknowns, with a corresponding computing time of $62$ seconds.

\begin{table}
\caption{Total number of unknowns, time to solution (seconds) and relative error on the computed $S_{ij}$ with respect to the reference solution for edge finite elements of degree $1$ and $2$ on different meshes.}
\label{tab:degree12}
\centering
$ \begin{array}{rrr} 
\multicolumn{1}{l}{\text{Degree 1}}   &  &  \\
\toprule
\text {\# unknowns} & \text{time (s)} & \text{error} \\ 
\midrule
2\,373\,214 & 22  & 0.384 \\ 
8\,513\,191 & 53  & 0.184 \\ 
21\,146\,710 & 130  & 0.117 \\
42\,538\,268 & 268  & 0.083\\
73\,889\,953 & 519  & 0.068\\
\bottomrule
\end{array}$
\qquad
$ \begin{array}{rrr} 
\multicolumn{1}{l}{\text{Degree 2}}   &  &  \\
\toprule
\text {\# unknowns} & \text{time (s)} & \text{error} \\ 
\midrule
1\,508\,916 & 39  & 0.243\\
5\,181\,678 & 62  & 0.099\\
12\,693\,924 &122  & 0.057\\
26\,896\,130 & 236  & 0.036\\
45\,781\,986 & 396  & 0.019\\
\bottomrule
\end{array}$
\end{table}

\newpage

\section{Conclusion}

This work shows the benefits of using a discretization of the time-harmonic Maxwell's equations based on high order edge finite elements  coupled with a parallel domain decomposition preconditioner for the simulation of a microwave imaging system.
In such complex systems, accuracy and computing speed are of paramount importance, especially for the application considered here of brain stroke monitoring.

Ongoing work consists in incorporating high order methods in the inversion tool that we are developing in the context of this application in brain imaging, for which promising results have already been obtained with edge finite elements of lowest order for the reconstruction from synthetic data of a numerical brain model.

We are also now in a position to test our inversion algorithm on various data sets acquired by the measurement system prototype of EMTensor.

From the numerical point of view, promising techniques are available that will allow us to speed up the solution of the inverse problem. First, recycling and block methods can be very helpful in such a context. The inverse problem is solved by a local optimization algorithm which consists in solving a sequence of slowly-varying linear systems, and a recycling algorithm such as GCRO-DR (Generalized Conjugate Residual method with inner Orthogonalization and Deflated Restarting)~\cite{Parks:2006:RKS} can significantly reduce the total number of iterations over all linear systems, by recycling the Krylov subspace from one linear system solve to the next. Moreover, each iteration in the inversion loop corresponds to solving a linear system with multiple right-hand sides available simultaneously, with one right-hand side per transmitting antenna. Each direct problem with multiple right-hand sides can thus be solved efficiently by block methods such as Block GMRES, or by combining block and recycling strategies in a Block GCRO-DR algorithm. Block methods provide higher arithmetic intensity and better convergence.

Finally, choosing a suitable coarse space for the design of a scalable two-level preconditioner for Maxwell's equations is still an open problem. Indeed, enriching the one-level preconditioner presented here with an efficient two-level preconditioner would lead to better convergence when using many subdomains, resulting in a highly scalable parallel solver.

\bibliographystyle{wileyj}
\bibliography{paperEMF}

\begin{thebibliography}{10}
\providecommand{\url}[1]{\texttt{#1}}
\providecommand{\urlprefix}{URL }
\expandafter\ifx\csname urlstyle\endcsname\relax
  \providecommand{\doi}[1]{doi:\discretionary{}{}{}#1}\else
  \providecommand{\doi}{doi:\discretionary{}{}{}\begingroup
  \urlstyle{rm}\Url}\fi

\bibitem{Semenov:2008:MTB}
Semenov SY, Corfield DR. Microwave tomography for brain imaging: feasibility
  assessment for stroke detection. \emph{International Journal of Antennas and
  Propagation}  2008; .

\bibitem{Persson:2014:MBS}
Mikael P, Andreas F, et~al. Microwave-based stroke diagnosis making global
  prehospital thrombolytic treatment possible. \emph{IEEE Transactions on
  Biomedical Engineering}  2014; .

\bibitem{Semenov:2014:ETB}
Semenov S, Seiser B, Stoegmann E, Auff E. Electromagnetic tomography for brain
  imaging: from virtual to human brain. \emph{2014 IEEE Conference on Antenna
  Measurements \& Applications (CAMA)}, 2014.

\bibitem{Nedelec:1980:MFE}
N{{\'e}}d{{\'e}}lec JC. Mixed finite elements in {${\bf R}^{3}$}. \emph{Numer.
  Math.}  1980; \textbf{35}(3):315--341, \doi{10.1007/BF01396415}.

\bibitem{Rapetti:2007:HOE}
Rapetti F. High order edge elements on simplicial meshes. \emph{M2AN Math.
  Model. Numer. Anal.}  2007; \textbf{41}(6):1001--1020,
  \doi{10.1051/m2an:2007049}.

\bibitem{RapBos:2009:WFH}
Rapetti F, Bossavit A. Whitney forms of higher degree. \emph{SIAM J. Numer.
  Anal.}  2009; \textbf{47}(3):2369--2386, \doi{10.1137/070705489}.

\bibitem{BDHR:2016:OSP}
Bonazzoli M, Dolean V, Hecht F, Rapetti F. Overlapping {S}chwarz
  preconditioners for high order edge finite elements: application to the
  time-harmonic {M}axwell's equations 2016. Preprint {HAL},
  https://hal.archives-ouvertes.fr/hal-01298938.

\bibitem{BonRap:2015:duality}
Bonazzoli M, Rapetti F. High-order finite elements in numerical
  electromagnetism: degrees of freedom and generators in duality.
  \emph{Numerical Algorithms}  2016; :1--26\doi{10.1007/s11075-016-0141-8}.

\bibitem{Hecht:2012:NDF}
Hecht F. New development in {F}ree{F}em++. \emph{J. Numer. Math.}  2012;
  \textbf{20}(3-4):251--265.

\bibitem{Dolean:2015:IDD}
Dolean V, Jolivet P, Nataf F. \emph{An Introduction to Domain Decomposition
  Methods: algorithms, theory and parallel implementation}. {SIAM}, 2015.

\bibitem{CaiSar:1999:RAS}
Cai XC, Sarkis M. A restricted additive {S}chwarz preconditioner for general
  sparse linear systems. \emph{SIAM J. Sci. Comput.}  1999;
  \textbf{21}(2):792--797 (electronic), \doi{10.1137/S106482759732678X}.

\bibitem{DolGanGer:2009:OSM}
Dolean V, Gander MJ, Gerardo-Giorda L. Optimized {S}chwarz methods for
  {M}axwell's equations. \emph{SIAM J. Sci. Comput.}  2009;
  \textbf{31}(3):2193--2213, \doi{10.1137/080728536}.

\bibitem{JolHecNat:2013:hpddm}
Jolivet P, Hecht F, Nataf F, Prud'Homme C. Scalable domain decomposition
  preconditioners for heterogeneous elliptic problems. \emph{Proc. of the Int.
  Conference on High Performance Computing, Networking, Storage and Analysis},
  IEEE, 2013; 1--11.

\bibitem{Parks:2006:RKS}
Parks ML, {D}e Sturler E, Mackey G, Johnson DD, Maiti S. {Recycling {K}rylov
  Subspaces for Sequences of Linear Systems}. \emph{SIAM Journal on Scientific
  Computing}  2006; \textbf{28}(5):1651--1674.

\end{thebibliography}

%
%
%

\end{document}